\documentclass[prd,aps,twocolumn]{revtex4}
\usepackage{graphicx,url}

\def\lsim{\raise0.3ex\hbox{$\;<$\kern-0.75em\raise-1.1ex\hbox{$\sim\;$}}}
\def\gsim{\raise0.3ex\hbox{$\;>$\kern-0.75em\raise-1.1ex\hbox{$\sim\;$}}}
\def\eps{\varepsilon}

\begin{document}

\title{Signatures of a two million year old supernova in the spectra of cosmic ray protons, antiprotons and positrons}

\author{M.~Kachelrie\ss$^{1}$}
\author{A.~Neronov$^{2}$}
\author{D.~V.~Semikoz$^{3}$}
\affiliation{$^1$Institutt for fysikk, NTNU, Trondheim, Norway}
\affiliation{$^2$Astronomy Department, University of Geneva,
Ch. d'Ecogia 16, Versoix, 1290, Switzerland}
\affiliation{$^3$Astroparticules et Cosmologie, 10 rue Alice Domon et Leonie Duquet, 
F-75205 Paris Cedex 13, France}

\begin{abstract}
The locally observed cosmic ray spectrum has several puzzling features, such as the excess of positrons and antiprotons above $\sim 20$\,GeV and the discrepancy in the slopes of the spectra of cosmic ray protons and heavier nuclei in the TeV--PeV energy range. We show that these features are consistently explained by a nearby source which was active $\sim 2$\,Myr ago and has injected  $ (2-3)\times 10^{50}$\,erg in cosmic rays. The transient nature of the source and its overall energy budget point to the supernova origin of this local cosmic ray source. The age of the supernova suggests that the local cosmic ray injection was produced by the same supernova  that has deposited $^{60}$Fe isotopes in the deep ocean crust.
\end{abstract}
\maketitle

\noindent
\textit{Introduction.}
Cosmic rays (CR) with energies at least up to $10^{15}$\,eV are thought to be a by-product of the final stages of stellar evolution~\cite{berezinsky_book,blasi_review,higdon}. The two main possibilities for the  acceleration sites of CRs are  individual supernovae (gamma-ray bursts, supernova remnants and pulsar wind nebulae) \cite{berezinsky_book,blasi_review} and superbubbles \cite{higdon} hosting large numbers of supernovae (SN) and  their progenitors, high-mass stars.

The direct identification of CR sources which would allow the discrimination between these two possibilities is difficult, because the turbulent Galactic  magnetic field (GMF) randomises the CR trajectories and leads to an almost isotropic CR intensity. Moreover, locally detected CRs are accumulated from a large number of sources which were active over the time scale $\tau_{\rm esc}\sim 10-30$\,Myr on which CRs (of the energy $E\sim 10$\,GeV) escape from the Galaxy. The superposition of the signals from a large number of sources erases possible signatures of individual sources.

The local CR flux might still have some "memory" of the individual sources composing it, because of the discrete and stochastic nature of the sources (be it SNe or superbubbles). The subset of near and recent CR sources could produce small features in the CR spectrum or create anisotropies~\cite{erlykin97,BA11,erlykin13,bernard13}. The identification of such features could potentially provide a possibility for the identification of the CR sources  and for the measurement of their characteristics.

In what follows we show that the known differences in the slopes between  CR protons and nuclei \cite{CREAM,CREAM_nuclei, dataKG},  puzzling features in the spectra of positrons \cite{PAMELA_positrons,PAMELA_positrons1,AMS02_positrons,AMS02_positrons1} and  
antiprotons \cite{PAMELA_antiprotons,AMS02_antiprotons} could be self-consistently explained  by a single nearby, recent CR source. We are able to deduce the characteristics of the source from the details of the spectra of these CR flux components.  In particular, the hard spectra of antiprotons and positrons above $\sim 20$\,GeV and the soft spectrum of CR protons (compared to the spectra of heavy nuclei) can be explained by a source which has injected $\sim 10^{50}$\,erg in CRs in a transient event which occurred  $\sim 2$\,million years ago. The source is located at a distance of (several) hundred parsecs along the local GMF direction. The transient nature of the event, its overall energy budget and the spectral characteristics of the injected CRs are consistent with a single SN and inconsistent with a superbubble as source.

%%%%%%%%%%%%%%%%%%%%%%%%%%%%%%%%%%%%%%%%%%%%%%%%%%%
\vskip0.2cm
\noindent
\textit{Contribution of a local source to the proton spectrum.}
%%%%%%%%%%%%%%%%%%%%%%%%%%%%%%%%%%%%%%%%%%%%%%%%%%%
Cosmic rays injected by a single source  $T$~years ago fill a region of the size 
$d_{\rm \|,\perp}\sim (D_{\|,\perp}T)^{1/2}$  in the interstellar medium (ISM). Here, $D_\|$ 
and $D_{\perp}$ are the energy dependent components of the diffusion tensor
parallel and perpendicular to the local GMF direction~\cite{casse,GKSS12,Giacinti:2012ar}. If the total injected energy ${\cal E}_{\rm tot}$ is high enough, the source could produce a significant increase in the overall CR flux detectable by observers situated inside the  region filled with CR. In the particular case of locally detected TeV CRs, the source contribution to the flux, $F\propto {\cal E}_{\rm tot}/(d_{\|}d_\perp^2)\propto {\cal E}_{\rm tot}/T^{3/2}$ could be comparable to the locally observed CR flux if ${\cal E}_{\rm tot}/T^3\sim 10^{50}\mbox{ erg}/(1\mbox{ Myr}^{3/2})\sim 10^{52}\mbox{ erg}/(10\mbox{ Myr})^{3/2}$. Thus, both a SN which occurred a million years ago and has injected $\sim 10^{50}$\,erg in CRs and a superbubble which has injected $\sim 10^{52}$\,erg over the last 10\,Myr can produce distortions in the local CR spectrum. 

Cosmic rays spread faster along the direction of the GMF, $D_\|\gg D_{\perp}$. The transient enhancement of the CR flux caused by a local source could be particularly strong if the source and the observer lie close to the same magnetic field line.  We model such a flux enhancement numerically using the code developed and tested in 
Ref.~\cite{GKSS12}.   The code follows the trajectories of individual CR particles through the GMF model of Jansson-Farrar~\cite{farrar}, starting from the moment of instantaneous injection in a single point by a transient CR source. The turbulent part of the field is chosen to follow isotropic  Kolmogorov turbulence with the maximal length of the fluctuations  $L_{\max}=25$\,pc and  the strength normalised to reproduce the observed 
B/C ratio, as discussed in~\cite{GKS2014,GKS2015}. 
The calculation of the trajectories of individual CRs in the GMF allows us  
to include a detailed model for the regular and the turbulent component
of the GMF.
  We record the path length of CRs spent in a 50\,pc sphere around the Earth, which can be converted to the local CR flux at given time interval.

We are interested in the case of a relatively young, $T\lsim {\rm few}$\,Myr, and nearby source,  $d_{\rm source}\sim {\rm few}\times 100$\,pc, and 
CRs  energies in the range 100\,GeV--100\,TeV. The spread of the CRs of such energy on Myr time scale is strongly anisotropic.
A strong enhancement of the CR flux occurs if the source and the observer are connected by a magnetic field line. In this case, the contribution of a single source can dominate the observed total CR intensity at the Earth.  

\begin{figure}
\includegraphics[angle=0,width=\columnwidth]{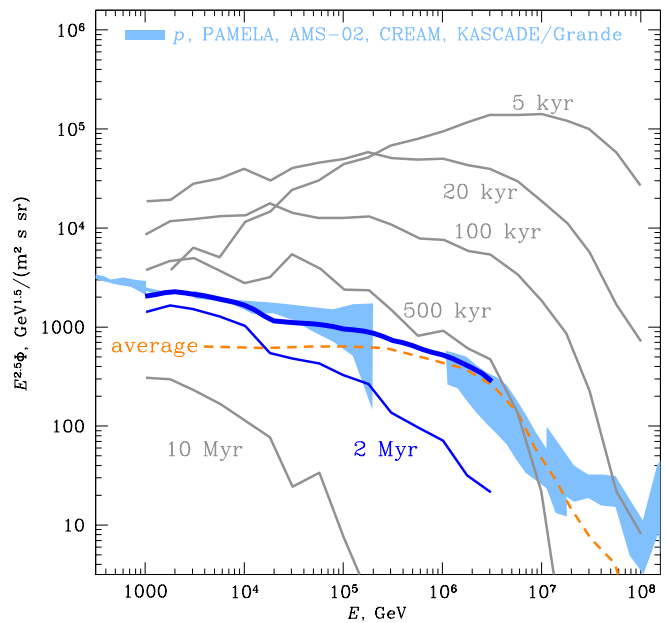}
\caption{Proton flux of the local source at different times. The average Galactic proton flux is shown as a thin orange line, the measured spectra of protons (light blue)  from PAMELA~\cite{PAMELA}, CREAM~\cite{CREAM}, KASCADE and KASCADE-Grande \cite{dataKG} as a band including experimental uncertainties. The sum of the average flux and the 2~Myr old source is shown by the blue thick line.  %The uncertainty of the average flux could be estimated from the uncertainty of the overall flux measurement in the PeV band.
}
\label{Pflux}
\end{figure}

Figure~\ref{Pflux} shows an example of such a situation calculated for a source at the distance 300~pc which has injected CRs with spectrum $dN/dE\propto E^{-\gamma_{\rm p,inj}}$, $\gamma_{\rm p,inj}=2.2$, and total injection energy ${\cal E}_{\rm tot}=2.5\times 10^{50}$\,erg. The source is placed at a GMF line passing within 50\,pc from the Solar system.
For $E>10$\,TeV, we calculate the CR trajectories up to 30\,Myr, i.e.\ 
sufficiently long to observe the exponential cutoff in the flux due to 
CR escape. At any given energy, we find that the observed flux $F$ at Earth 
as function of time rises, 
then drops as a power-law $F(t) = F_{\max} (t_0/t)^{\alpha (E)}$ up to the 
(energy-dependent) escape time  and finally is exponentially suppressed as 
$F(t) = F_{\max}(t_0/t)^{\alpha (E)}\exp(-t/\tau_{\rm esc})$.
In the energy range 1--10\,TeV,  we are only able to calculate trajectories 
up to 300\,kyr. We extrapolate them to later times using the power-law 
with the slope $\alpha(E)$ derived from direct simulations in the
energy range 10\,TeV--1\,PeV. 
Note that the fluctuations visible especially
at large times are due to the relatively small number of CR trajectories used.

%%%%%%%%%%%%%%%%%%%%%%%%%%%%%%%%%%%%%%%%%%%%%%%%%%%
%%%%%%%%%%%%%%%%%%%%%%%%%%%%%%%%%%%%%%%%%%%%%%%%%%%

From Fig.~\ref{Pflux} one can see that CRs with energies above 100\,TeV reach the Earth already 5\,kyr after the injection. If the source is able to accelerate CRs to energies above 10\,PeV, their flux is suppressed already after 5\,kyr, because of the fast escape from the Galactic disk. The escape induced flux suppression progresses towards lower energies with the increase of the source age. Below the high-energy cut-off, the slope of the spectrum softens and reaches  the observed value $\tilde\gamma_p\sim 2.7-2.8$ after 2\,Myr.

The observed slopes of the spectra of the heavy nuclei component of the CR flux, $\gamma_{N}\simeq 2.5$, are systematically harder than the slope of the proton spectrum in the TeV--PeV range \cite{CREAM_nuclei,CREAM}. This harder slope of the nuclear component of the CR flux consistently explains the shapes of the knees in the spectra of individual groups of nuclei within
the escape model~\cite{GKS2014,GKS2015}.  The same slope of the average spectrum of 0.1--10\,TeV protons/nuclei  in the Galaxy  is deduced from a combination of gamma-ray and IceCube neutrino data~\cite{NST2013,NS2014,neronov15}. 

Assuming that the average Galactic CR proton flux at the Earth also has the slope $\gamma_p\simeq 2.5$  and that it dominates the observed CR flux in the knee energy range $E\sim 1$--10\,PeV (shown as ``average'' in Fig.~\ref{Pflux}, see Ref.~\cite{GKS2014}) one finds that the local source and the average Galactic contributions to the overall CR proton fluxes are comparable in the energy range 3--30\,TeV. 
The uncertainty of the average flux is given by the uncertainty of measurements around the knee, cf.\ with the width of the blue band at PeV.
The  contribution of the local source with its softer spectrum explains the discrepancy between the slopes of the proton and heavy nuclei components of the TeV--PeV CR spectrum. In general, the local source gives also a contribution to the spectra of heavier nuclei. If the elemental abundance of the local source CRs is identical to the overall measured CR abundance, the contribution of the source to the heavy nuclei spectra in the $E>1$~TeV range is sub-dominant, because of the higher normalisation of the average Galactic component of the heavy nuclei fluxes.

%%%%%%%%%%%%%%%%%%%%%%%%%%%%%%%%%%%%%%%%%%%%%%%%%%%
\vskip0.2cm
\noindent
\textit{Positron excess from the local CR proton source.}
%%%%%%%%%%%%%%%%%%%%%%%%%%%%%%%%%%%%%%%%%%%%%%%%%%%
Our suggestion that the softer slope of the TeV-PeV proton 
CR spectrum is caused by a local source can be tested via the 
identification of complementary signatures in the spectra of secondary 
particles---positrons and antiprotons----produced in CR interactions in 
the ISM.

The spectrum of CR positrons is known to have an "excess" above 30\,GeV. 
This excess refers to  a deviation from reference models (as e.g.\ those of
Ref.~\cite{gal}) which assume that positrons are
solely secondary particles produced during the propagation of time independent CR proton 
and nuclei flux through the ISM. The presence of this excess is usually 
considered as an indication for the existence of a source of positrons 
in the local Galaxy.  Source candidates under discussion are 
nearby pulsars~\cite{blasi_serpico}, young ($\sim 10^4$~yr) supernova remnants~\cite{blasi09} 
and dark matter annihilations or decays~\cite{DMreview}. 

A characteristic feature of secondary production in hadronic interactions
is that the slope of the energy spectrum of secondaries is very
close to the one of the parent protons, since scaling violations are
small except close to mass thresholds. Thus, the presence of a local 
source of CR protons should reveal 
itself through an associated component in the CR positron spectrum. 
The average energy fraction transferred to  positrons produced in a CR 
interaction is $\langle z_{e^+} \rangle \simeq 3\%$ for $\gamma=2.2$,
so that positrons with energies in the 30--300\,GeV range are produced by 
protons with energies $\gsim 1$--10\,TeV.  
The flux of this local positron component is a function of time. It grows as
the fraction of protons interacting with the ISM increases, but keeps 
approximately the shape of the parent proton distribution. Our calculation
of the CR trajectories emitted from the local source presented in the 
previous section  allows us to determine the average grammage $X$ traversed 
by the  CRs since the moment of injection. For energies $E\lsim 10^{14}$\,eV,
the grammage is nearly energy independent, $X\simeq 0.3$\,g/cm$^2$, 
for a source of the age $T=2$\,Myr.

The produced positrons diffuse and spread over larger and larger distances, 
softening thereby their energy  spectrum. The process of diffusion and
the resulting softening of the spectral slope is identical for positrons
and protons, if the age of the local source 
is small enough that energy losses of the positrons can be neglected. 
Thus, not only the injection, but also the propagated spectra of protons 
and positrons from the local source have nearly identical slopes at 
any moment of time. In particular, this implies that at present
$\gamma_{e^+}\simeq \tilde\gamma_p\simeq 2.7-2.8$.

%%%%%%%%%%%%%%%%%%%%%%%%%%%%%%%%%%%%%%%%%%%%%%%%%%%
\begin{figure}
\includegraphics[width=\columnwidth]{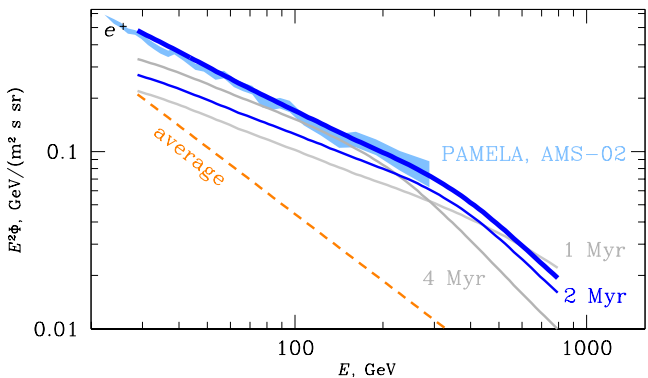}
\caption{Spectum of positrons from the local source for the age 2 Myr (thin blue curve), 1 Myr (light grey) and 4~Myr (darker grey), compared to the measured spectra of positrons \cite{PAMELA_positrons,PAMELA_positrons1,AMS02_positrons,AMS02_positrons1}. The dashed orange line shows an estimate of the average Galactic positron flux~\cite{donato09}.} 
\label{fig:spectra}
\end{figure}
%%%%%%%%%%%%%%%%%%%%%%%%%%%%%%%%%%%%%%%%%%%%%%%%%%%

Figure~\ref{fig:spectra} shows the measured positron spectrum in the 30--300\,GeV energy range~\cite{PAMELA_positrons,PAMELA_positrons1,AMS02_positrons,AMS02_positrons1} together with our calculation of the positron flux from the local source.
For the calculation of the hadronic production cross sections, we have been
employing QGSJET-II-04~\cite{QGS} in the modified version presented in~\cite{antiproton_new}. 
Since we have fixed both the contribution of the local source to the proton flux and the grammage in Sec.~1, the normalisation of the shown positron flux is a prediction. Both the normalisation and the slope $\gamma_{e^+}\simeq \tilde \gamma_p$ agree well with the experimental data.

An additional suppression of the positron flux may occur due to synchrotron 
and inverse Compton energy losses.  The synchrotron and inverse Compton 
cooling rate is 
%$t_{s,IC}= 2\left[U/(0.5\mbox{ eV/cm}^3)\right]^{-1}\left[E_{e^+}/(300\mbox{ GeV})\right]^{-1}$~Myr 
$t_{\rm s,IC}^{-1}= 0.5 [U/(0.5\mbox{ eV/cm}^3)][E_{e^+}/(300\mbox{ GeV})]$Myr$^{-1}$,
where $U$ is the combined energy density of radiation and the magnetic field, $U=0.5\left[B/(4\ \mu\mbox{G})\right]^2$~eV/cm$^3$. The synchrotron and inverse Compton cooling softens the positron spectrum. Contrary to the primary CRs, which are injected instantaneously from a point source, positrons are continuously produced at a constant rate.
This leads to the formation of a cooling break from $\gamma_{e^+}$ to $\gamma_{e^+}+1$ in the positron spectrum. The break energy decreases with time as shown 
schematically in Fig.~\ref{fig:spectra}.  
The non-observation of such a softening limits the age of the local source to  $T\lesssim 4$~Myr. 

Note that the association of the observed positron excess with a local source also implies a lower limit on the source age, $T\gtrsim 2$\,Myr. Otherwise, if the source would be much more recent, CRs would not have enough time to produce the observed excess positron flux. 

Overall, the hypothesis of the local source contribution to the CR proton spectrum passes the positron self-consistency check and provides an explanation to the observed excess in the positron spectrum.

%%%%%%%%%%%%%%%%%%%%%%%%%%%%%%%%%%%%%%%%%%%%%%%%%%%
\vskip0.2cm
\noindent
\textit{Antiproton flux from the local CR source.}
%%%%%%%%%%%%%%%%%%%%%%%%%%%%%%%%%%%%%%%%%%%%%%%%%%%
Interactions of CR protons with the ISM also produce antiprotons.
The relative size of the antiproton and positron fluxes is in the
regime of negligible positron energy losses completely determined
by ratio of the corresponding $Z$-factors, or approximately by the
ratio of the spectrally averaged  energy fraction $\langle z_{i}\rangle$
transferred to  antiprotons and positrons~\cite{antiproton_new}.
We use again the modified version of QGSJET-II-04~\cite{antiproton_new} for 
the calculation of the spectrum of secondary antiprotons, keeping all
parameters fixed to the ones used for positrons. The resulting
spectrum of secondary antiprotons from the local source is shown in
 Fig.~\ref{fig:spectra} by the blue line. 
The  light-blue shaded range shows the $\pm 50\%$ uncertainty band 
attributed to the uncertainty of the antiproton production cross-section
in this energy range~\cite{antiproton_uncertainty,antiproton_uncertainty1,antiproton_new}. 

From this figure one can see that the flux of the local source antiprotons constitutes a significant fraction of the overall CR antiproton flux. 
The local source contribution to the antiproton to proton ratio is at the level $\sim 10^{-4}$, which is comparable to the measured ratio \cite{PAMELA_antiprotons,AMS02_antiprotons}.
Note that the very preliminary AMS-02 antiproton measurements reported 
in~\cite{AMS02_antiprotons} are consistent with the published PAMELA 
results~\cite{PAMELA_antiprotons} up to 180\,GeV.
The apparent independence of the $\bar p/p$ ratio with energy is at tension with the naive expectation that in the steady state model of secondary antiproton production by CR interactions in the ISM  the antiproton spectrum should be softer than the proton spectrum, so that the ratio should have decreasing trend with the increase of energy \cite{DM-AMS02}. 
The presence of the single source contribution to the antiproton flux removes this tension. The energy dependence of the $\bar p/p$ ratio for the single source and for the average Galactic CR flux are different. Antiprotons originating from the source are injected by the parent protons which have spent the same time in the ISM, independently of their energy. For this component, the $\bar p/p$ ratio is rather slowly rising for energies well above the antiproton production threshold. The resulting independence of the $\bar p/p$ ratio 
with energy is therefore a falsifiable prediction of our model.
%
%%%%%%%%%%%%%%%%%%%%%%%%%%%%%%%%%%%%%%%%%%%%%%%%%%%
\begin{figure}
\includegraphics[width=\columnwidth]{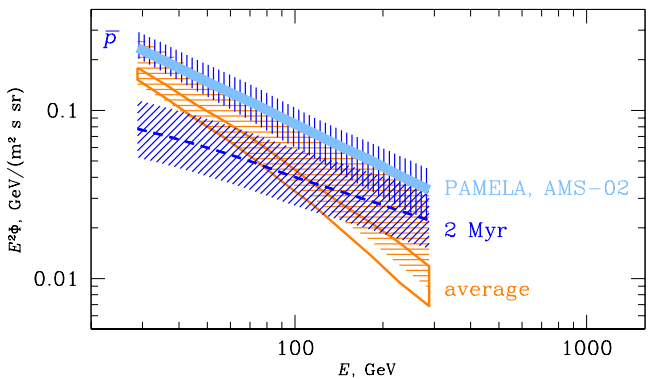}
\caption{Spectrum of antiprotons from the local source (dashed blue line) compared to the measurements from~\cite{PAMELA_antiprotons,AMS02_antiprotons} (thick light blue line).  The orange box shows an estimate of the average Galactic antiproton flux from~\cite{donato09}; the horizontally orange hatched range shows an alternative calculation  from~\cite{DM-AMS02}. The blue hatched ranges show the uncertainty of the model calculation (inclined hatching) and for the sum of the local source and average antiproton fluxes (vertical hatching).}
\label{fig:ratio}
\end{figure}
%%%%%%%%%%%%%%%%%%%%%%%%%%%%%%%%%%%%%%%%%%%%%%%%%%%
%
Overall,  the presence of the local source component in the antiproton flux is consistent and possibly even favoured by the $\bar p/p$ ratio data.

%%%%%%%%%%%%%%%%%%%%%%%%%%%%%%%%%%%%%%%%%%%%%%%%%%%
\vskip0.2cm
\noindent
\textit{Supernova nature of the local CR source.}
%%%%%%%%%%%%%%%%%%%%%%%%%%%%%%%%%%%%%%%%%%%%%%%%%%%
The combination of the positron, antiproton and proton signatures of a local CR source provides a possibility to constrain its parameters.  The source should have the age of $T\sim 2$\,Myr. An older source would 
be inconsistent with the absence of radiative cooling in the positron 
spectrum. A younger source would fail to produce sufficient amount of 
antimatter. A younger source would provide also a harder feature in the 
proton spectrum in the TeV range, while for an older source the source 
contribution would be too soft to be noticeable in the TeV band.  

The overall energy injected in CRs should be at the level of 
${\cal E}_{\rm tot}\sim 10^{50}$~erg, for a wide range of source distances. 
The energy density of high-energy particles inside a region in the ISM 
filled with CRs injected by the source is nearly uniform. The region spans 
a $\sim 100$\,pc wide tube of kpc-scale length the along the local GMF direction. The only condition for the detectability of the  CR flux from the local source is that the Solar system is situated inside 
this region.  

The source has to be transient, in the sense that it could not remain active 
all through the last million years. Otherwise the source would strongly 
contribute to the CR flux from more recent injection times. For example, 
the  contribution from the last $0.1$\,Myr period would have the spectrum
shown by the orange curve in Fig. \ref{Pflux} multiplied by 0.1\,Myr/1\,${\rm Myr}=0.1$. This contribution would be comparable to the 2\,Myr old contribution. However, the energy spectrum of this younger contribution 
is harder, with the slope $\gamma\simeq 2.5$. The presence of such a 
contribution would erase the effect of softening and the spectrum of 
TeV--PeV protons would have the same slope as the spectrum of heavy nuclei. 

The transient nature of the source and the overall injected energy 
rule out the possibility that the local source is a superbubble blown 
by massive star formation. The typical lifetime of superbubbles is 
in the $10^7$--$10^8$\,yr range determined by the lifetime of massive stars. 
A young superbubble formed a million years ago would be still active today.

The only plausible model of the transient local CR source is that of a SN. 
The average rate of SN explosions in the Milky Way disk volume $V_{\rm disk}\simeq 100$--300\,kpc$^3$ is ${\cal R}_{SN} \simeq (1$--$3)\times 10^{-2}$\,yr$^{-1}$, 
or one SN per (0.3--3)$\times 10^4$\,yr per kpc$^3$. This means 
that one could reasonably expect that about one SN has exploded within 
the last million years within a 100\,pc wide, kpc long filament 
directed along the GMF line going through the Solar system.

%%%%%%%%%%%%%%%%%%%%%%%%%%%%%%%%%%%%%%%%%%%%%%%%%%%
\vskip0.2cm
\noindent
\textit{Discussion.}
%%%%%%%%%%%%%%%%%%%%%%%%%%%%%%%%%%%%%%%%%%%%%%%%%%%
Our analysis has shown that several features in the CR spectrum which 
appear puzzling within the standard  Galactic CR injection/propagation models find a natural 
explanation by the presence of a local CR source. In particular, we have 
shown that a $\sim 2$\,Myr old source which injected $\sim 10^{50}$\,erg in 
CRs with energies up to at least 30\,TeV can consistently explain 
the difference in the slopes of the proton and heavy nuclei spectra 
in the TeV--PeV  energy range,  give an additional contribution to the antiproton spectrum in agreement with recent AMS-02 data and predict the correct amplitude and slope for the positron spectrum 
in the 30--300\,GeV energy range. 
The local source also gives rise to an excess anisotropy of the CR spectrum in the 1--100\,TeV energy range, which for first time explains the anisotropy data in this energy range~\cite{savchenko15}. Note that the combination of the anisotropy and the positron data constrain the source age and disfavor suggestions of younger sources in denser environments as in Ref.~\cite{Fujita:2009wk}.

The scenario of a 2\,Myr old local source contributing to the 1--100\,TeV CR spectrum could be tested with future AMS-02 and ISS-CREAM measurements of heavy nuclei CR fluxes. Such measurements could detect the imprint of the local source on the primary-to-secondary CR nuclei ratios. In particular, the energy independent grammage ($\simeq 0.3$~g/cm$^2$) traversed by Carbon nuclei should result in the presence of an energy-independent component of the B/C ratio in the 1-100~TeV energy band. The 2\,Myr time scale is also comparable to the decay time of $^{10}$Be nuclei. This may leave an imprint in the $^{10}$Be/$^{9}$Be ratio in the same energy range.

The presence of a nearby SN explosion was previously noticed in a 
completely different type of data on abundance of isotopes on 
Earth~\cite{Ellis,knie99,benitez,Ellis1}. These data suggest that an episode 
of deposition of $^{60}$Fe isotopes in the million years old deep ocean crust 
was produced by the passage of an expanding shell of a 2\,Myr old supernova
remnant through the Solar System. The consistency of the SN age and 
distance estimate from the deep ocean sediment data  with those found from 
CR data suggests it is one and the same supernova which
 is responsible for the CR injection and the isotope deposition on the Earth.

%\end{document}

\newpage

\appendix

\section{Supplementary material}

\subsection{Trajectory approach}

In the standard approach to Galactic CR propagation, the $N$-particle phase 
space distribution of CRs is approximated as a macroscopic fluid.  
Solving then the diffusion equation including CR 
interactions permits to model self-consistently  primary and secondary 
fluxes. However, the diffusion approach has several disadvantages. First 
of all, the diffusion approximation itself breaks down in several interesting 
regimes~\cite{1,2,3}. Moreover, the diffusion tensor  is an external input 
in this approach. It is only loosely connected to fundamental properties 
of the Galactic magnetic field (GMF). In this work, we apply 
therefore the trajectory approach where one calculates the path of individual 
CRs solving the equations of motion of particles propagating in the GMF. 
We use the code described and tested 
in~\cite{0}, while we use for the regular and the turbulent component of the 
GMF the parameters described in \cite{4}. As we showed previously in 
\cite{3,4}, the resulting CR fluxes reproduce over a wide range of energies 
all available experimental  data for individual groups of CR nuclei.

\begin{figure}
\includegraphics[width=0.7\columnwidth,angle=270]{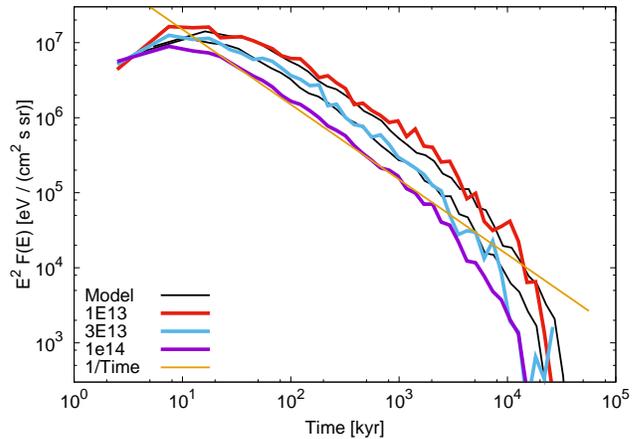}
\caption{The number density $n(E,t)$ of CRs at the position of the Earth 
as function of time for three 
different energies; also shown are the densities rescaled using 
$\tilde t= (E_0/E)^{1/3}$ (grey lines) and the $1/t$ behavior expected for 
$d=2$ diffusion.}
\label{fig:s1}
\end{figure}

Our numerical results for the propagation of CRs emitted by a bursting 
source can be summarized as follows. In an initial period, CRs propagate 
essentially one-dimensional, filling a filament-like structure around the
magnetic field line going through the source. 
After this rather short initial period, CRs start to diffuse also
perpendicular to the direction of the filament. In a realistic GMF model,
the two perpendicular components of the diffusion tensor differ,
$D_{\perp,1}\gg D_{\perp,2}$, and thus CRs diffuse mainly in the $\|$ and
the $\perp,1$ direction. In particular, we found for the case at hand in
the Jansson-Farrar  model~\cite{Sfarrar} the
values $\{D_\|,D_{\perp,1},D_{\perp,2}\}\sim \{10^{30},10^{28},10^{27}\}$\,cm$^2$/s
for CRs with energy $100$\,TeV. 
This behavior is illustrated in Fig.~\ref{fig:s1} which shows the
scaled CR density $n$ at the position of the Earth as function of time
for three different energies. Additionally, we used the relation 
$D(E)t\propto E^{1/3}t$ valid for 
Kolmogorov turbulence to rescale the $E_0=10^{14}$\,eV curve by
$t\to \tilde t=t (E_0/E)^{1/3}$ to $E=3\times 10^{13}$\,eV and to
$E=1\times 10^{13}$\,eV. The approximate agreement of the
two rescaled curves with the true result supports a Kolmogorov-like 
scaling of the density.
Moreover, it is visible that the density scales approximately as 
$n\propto 1/t$ for intermediates times,
10\,kyr$\lsim \tilde t\lsim 10$\,Myr. Such a scaling is 
characteristic for two-dimensional diffusion and reflects the observed
hierarchy in the eigenvalues of the diffusion tensor.

%%%%%%%%%%%%%%%%%%%%%%%%%%%%%%%%%%%%%%%%%%%%%%%%%%%
%%%%%%%%%%%%%%%%%%%%%%%%%%%%%%%%%%%%%%%%%%%%%%%%%%%

\subsection{Secondary fluxes}

The trajectories of charged CRs depend on their rigidity 
${\cal R}=E/(Ze)$. 
The sign difference of the charge of protons and 
antiprotons becomes important only on distances of a few coherence lengths 
to the source, while on larger scales the diffusion of protons and 
antiprotons of the same energy proceeds in the same way. In this paper we do not perform 
the full (computationally expansive) calculation of trajectories of the secondary particles. 
Instead, we use a scaling relation to describe the spread of the secondaries, taking into 
account the shift of energy of the secondaries compared to the primary particles. 

The density  $n_{\bar p}(E',r',t')$ of antiprotons 
which are produced  at the time $t$ by protons with the density  
$n_{p}(E,r,t)$ is given by
\begin{equation}
\label{eq:Green}
 n_{\bar p}(E',r',t')= \int d^d x \: G_{\bar p}(r',t';r,t) n_{p}(E,r,t) \,,
\end{equation}
where $G_{\bar p}(r',t';r,t)$ is the Green function describing the diffusion
of antiprotons. Our numerical results show that the CR propagation from 
a bursting source can be modeled by two-dimensional diffusion in the time 
and energy range of interest for CR secondary production. Thus the
proton density at the production time $t$ is given by
\begin{equation}
\label{eq:n}
 n(E,r,t)= \frac{Q(E)}{\pi^{d/2}  r_{\rm diff}^{d}} \,
 \exp\left[-r^2/ r_{\rm diff}^2\right] \,,
\end{equation}
where $d=2$, $r$ denotes the distance in the two-dimensional diffusion plane, 
$Q(E)$ the source spectrum and $r_{\rm diff}$ the effective 
diffusion distance, $r_{\rm diff}^2=4Dt$. The corresponding Green function $G$ 
has the same functional form, 
\begin{equation}
\label{eq:G}
 G_{\bar p}(r',t';r,t) = \frac{1}{\pi^{d/2}  r_{\rm diff}^{\prime d}} \,
 \exp\left[-(r-r')^2/ r_{\rm diff}^{\prime 2}\right] 
\end{equation}
with $r_{\rm diff}^{\prime 2}=4 D'(t'-t)$.
For simplicity, we assume a one-to-one relation between the energy
of protons and antiprotons, setting $E'=a^3E\sim 0.1E$, where $a\simeq 0.46$. This results in
the diffusion coefficient $D'=aD$ for Kolmogorov turbulence.

The integral in Eq.~(\ref{eq:Green}) is of the standard Gaussian type. In
the limit of small $r$ the exponent is  
$\exp\left[-r^2/ r_{\rm diff}^2\right]\sim 1$ and  the ratio $n_{\overline p}/n_p$ is
\begin{equation}
\label{R}
R_{\bar p} \equiv \frac{n_{\bar p}}{n_{p}} \simeq \frac{1}{x+a(1-x)} \,,
\end{equation}
where we introduced also $x=t'/t$.
In particular, we see that the ratio does not depend on energy and
satisfies as required $R_{\bar p}(a=1)=R_{\bar p}(x=1)=1$.
Integrating over the production time $t$, we find 
$\langle R_{\bar p}(x)\rangle =\ln(a)/(a-1)\simeq 1.43$ for $a\simeq 0.46$.
Thus, 
using for CR antiprotons the primary proton trajectories we underestimate the antiproton 
flux by $\sim 50$\%. 
Since this enhancement is energy independent, we 
can use the following scheme for the calculation of the antiproton flux.

In a first step, we calculate the proton flux at the position of the
Earth neglecting interactions. In the trajectory approach, we obtain 
the local CR flux at a given time interval recording the path length of CRs 
spent in a sphere around this point.
We record also the grammage $X$ these 
CRs crossed on their way to the Earth. Then we calculate the average 
grammage $\langle X\rangle =N^{-1}c\sum_{i=1}^N\int dt\,\rho(\vec x_i(t))$ 
summing up the gas density along the trajectories of the $N$ individual CRs.
We employ $n(z)=\rho/m_p=n_0\exp(-(z/z_{1/2})^2)$ as model for the gas 
distribution in the Galactic disk, with $z$ as the distance to the Galactic 
plane, $n_0=0.3/$cm$^3$ at the Solar position  and $z_{1/2}= 0.21$\,kpc~\cite{gas}. We set also $n=10^{-4}$g/cm$^3$ as minimum gas 
density up to the edge of the Milky Way at $|z|=10$\,kpc. 
Next we calculate the secondary antiprotons produced by CRs interacting
with gas using the modified version of QGSJET-II-04 introduced in~\cite{ap}. 
The antiproton flux is then given by
\begin{equation}
\label{Ip}
 I_{\bar p}(E_{\bar p}) = R_{\bar p}\eps_{\rm nuc}  X/m_p  \;
   Z_{\bar p}(E_{\bar p},\alpha) \: I_p(E),
\end{equation}
where the $Z$-factor is given by the inclusive
spectra of antiprotons $d\sigma_{\bar p}(E,z_{\bar p})/d z_{\bar p}$, 
$z_{\bar p}=E_{\bar p}/E$, 
\begin{equation}
\label{Z_spec}
 Z_{\bar p}(E_{\bar p},\alpha) = \int_0^1 d z \, z^{\alpha-1}\,
 \frac{d\sigma_{\bar p}(E_{\bar p}/z,z)}{d z}\,. 
\end{equation}
with $\alpha\simeq 2.6$ as the average slope of the proton spectrum.
The factor $R_{\bar p}\simeq 1.5$ accounts for the 
underestimate of the antiproton flux, while we include the effect of 
heavier nuclei in the interstellar medium using a nuclear enhancement
factor $\eps_{\rm nuc}\sim 1.2$.
Note that it is a virtue of our model that the slope of the injection
spectrum is the same for all CR nuclei. As a result, the use of a
nuclear enhancement factor is justified.

The same approach could be applied also to positrons, with a change 
$a\simeq 0.31$  to take into account their lower average energy. The 
above simple picture has to be modified at the highest energies (beyond the current 
measurements energy range) 
where continuous energy losses of positrons  become important. 
The first estimate of the corrections due to the continuous energy losses  
could be found from the 
solution of the 
diffusion equation in the Syrovatskii approach in $d=2$~\cite{b},
\begin{equation}
\label{eq:nburst}
 n(E,r)= \frac{Q(E)b(E_i)}{\pi b(E) r_{\rm diff}^{2}} \,
 \exp\left[-r^2/ r_{\rm diff}^2\right] \,,
\end{equation}
where $r_{\rm diff}^2=4\int_0^t dt\,D(E)$, $b(E)=-dE/dt=\beta E^2$ the energy loss rate and $E_i$ denotes 
the energy at injection~\cite{b}. The effective 
diffusion coefficient for this solution is  $D\sim D_\|$ as determined from our simulations.
For a fixed value of $\beta$, this determines the electron and positron 
spectra formed at time $t$. The total electron and positron fluxes are 
then obtained summing over the formation time. This approximation has to be verified 
by simulations of trajectories 
of secondary positrons taking into account continuous energy losses.  
Such simulations require however
a considerable increase of computing resources and are therefore postponed to
a future work.

\end{document}